\documentclass[10pt, conference]{IEEEtran}
\IEEEoverridecommandlockouts

\usepackage{cite}
\usepackage{amsmath,amssymb,amsfonts}
\usepackage{algorithmic}
\usepackage{graphicx}
\usepackage{textcomp}
\usepackage{xcolor}
\usepackage{booktabs}    
\usepackage{multirow}    
\usepackage{tabularx}    
\usepackage{array}       
\usepackage{xcolor}
\usepackage{graphicx}    

\def\BibTeX{{\rm B\kern-.05em{\sc i\kern-.025em b}\kern-.08em
    T\kern-.1667em\lower.7ex\hbox{E}\kern-.125emX}}

\makeatletter
\def\subsubsection{\@startsection{subsubsection}{3}{\parindent}{1ex plus 0.1ex minus 0.1ex}%
    {0.7ex plus .5ex minus 0ex}{\normalfont\normalsize\itshape}}%
\makeatother

\makeatletter
\newcommand{\linebreakand}{%
  \end{@IEEEauthorhalign}
  \hfill\mbox{}\par
  \mbox{\hspace{0.5\marginparwidth}} 
  \hfill\begin{@IEEEauthorhalign}
}
\makeatother
\begin{document}
\title{FSD50K-Solo: Automated Curation of Single-Source Sound Events \\
\thanks{This work was conducted during Ningyuan Yang’s internship at Bose and was supported fully by Bose Corporation. This work has been accepted for publication at the 2026 European Signal Processing Conference (EUSIPCO).}
}

\author{\IEEEauthorblockN{Ningyuan Yang}
\IEEEauthorblockA{\textit{Electrical \& Computer Engineering} \\
\textit{Stony Brook University}\\
Stony Brook, NY, USA \\
ningyuan.yang@stonybrook.edu}
\and
\IEEEauthorblockN{Sile Yin}
\IEEEauthorblockA{\textit{Research} \\
\textit{Bose Corporation}\\
Framingham, MA, USA \\
sile\_yin@bose.com}
\and
\IEEEauthorblockN{Li-Chia Yang}
\IEEEauthorblockA{\textit{Research} \\
\textit{Bose Corporation}\\
Framingham, MA, USA \\
richard\_yang@bose.com}
\linebreakand
\IEEEauthorblockN{Bryce Irvin}
\IEEEauthorblockA{\textit{Research} \\
\textit{Bose Corporation}\\
Framingham, MA, USA \\
bryce\_irvin@bose.com}
\and
\IEEEauthorblockN{Xiao Quan}
\IEEEauthorblockA{\textit{Research} \\
\textit{Bose Corporation}\\
Framingham, MA, USA \\
xiao\_quan@bose.com}
\and
\IEEEauthorblockN{Marko Stamenovic}
\IEEEauthorblockA{\textit{Research} \\
\textit{Bose Corporation}\\
Framingham, MA, USA \\
marko\_stamenovic@bose.com}
\and
\IEEEauthorblockN{Shuo Zhang}
\IEEEauthorblockA{\textit{Research} \\
\textit{Bose Corporation}\\
Framingham, MA, USA \\
shuo\_zhang@bose.com}
}
\maketitle
\begin{abstract}
High-quality training datasets are essential for the performance of neural networks. However, the audio domain still lacks a large-scale, strongly-labeled, and single-source sound event dataset. The FSD50K dataset, despite being relatively large and open, contains a considerable fraction of multi-source samples where background interference or overlapping events could limit the usefulness of the data. To address this challenge, we introduce a data curation framework designed for large-scale open audio corpora. Our approach leverages a generative diffusion model to synthesize clean single-class events to construct controlled noisy mixtures for supervision. We subsequently employ a pre-trained audio encoder coupled with a discriminative classifier to automatically identify and filter out multi-source samples. Experiments show that our framework achieves strong performance on a human expert-curated test set. Finally, we release FSD50K-Solo, a model-curated subset of FSD50K containing single-source audio samples identified by our method. Beyond FSD50K, our method establishes a scalable paradigm for curating open source audio corpora.
\end{abstract}

\section{Introduction}
\label{sec:intro}

Data-driven approaches, especially owing to the application of deep learning models, have achieved remarkable success across multiple audio applications this century~\cite{luo2018tasnet,dong2018musegan,irvin2024fullband,ma2025real,yang2026survey}.
Tasks such as sound event localization and detection (SELD/SED)~\cite{hu2025pseld, shashaank2023hiss}, audio captioning~\cite{labbe2024, mei2024}, target sound extraction (TSE)~\cite{veluri2023realtime, wakayama25_interspeech}, and, more recently, large audio-language models (LALMs)~\cite{tang2024salmonn,xu2025qwen25omnitechnicalreport, goel2025audioflamingo3advancing} all require large-scale labeled datasets to achieve optimal performance. 
However, the quality of datasets often becomes the bottleneck for further performance improvements, reliable evaluation, and reproducible research~\cite{10094617, li2025moredatacurationmatters}.

Several previous works have shown that many popular audio datasets suffer from weak labeling, and that improvements in dataset quality lead to better model performance. 
For example, Hershey et al.~\cite{shawn2021} found that adding strong temporal labels to a small portion of AudioSet~\cite{gemmeke2017audio} could significantly improve performance in audio event classification. 
Mei et al.~\cite{mei2024} also showed that using labels generated by large language models (LLMs) can improve performance on tasks such as text-based sound generation and audio-language retrieval.

FSD50K~\cite{fonseca2021} represents one of the largest open and accessible sound event datasets, making it a critical resource for the audio research community. 
However, the authors acknowledge several limitations, one of which is the overall weakness of the labels. Labels are provided at the clip level, with no further information about which sound events occur at a given time. For clips with only one sound event, which we term single-source samples, this is not an issue. However, clips with multiple types of sound events that may be sparsely distributed and/or overlapping result in noisy labels. 

Moreover, many audio machine learning training workflows use data perturbation pipelines at the model input to create synthetic mixtures whose constituent parts, or metadata about them, can be used as training targets for optimization. Ensuring that these constituent parts are single-source samples yields mixtures which are themselves strongly labeled, allowing for training without noise from spurious input or target data. 

This motivates our work to build a framework that robustly identifies all single-source samples in large-scale audio corpora. 
The framework employs a pre-trained audio encoder combined with a discriminative classifier trained on a small set of single-source reference data generated by a diffusion-based audio model. 
In this paper, we describe how the proposed system is designed to scale while leveraging minimal human labor, and utilize it to clean up a corpus, or corpora, that are orders of magnitude larger. We also showcase the framework by creating FSD50K-Solo, a subset of FSD50K containing only single-source audio clips.

In Section~\ref{sec:related_work} we review related work in audio data filtering and quality assessment.
In Section~\ref{sec:methods}, we detail our scalable approach for generating FSD50K-Solo\footnote{Dataset Link: https://doi.org/10.5281/zenodo.20403806}, the single-source subset of FSD50K curated by our method. Sections~\ref{sec:experiments} and~\ref{sec:results} showcase experiments demonstrating the effectiveness of our method. We release the complete clip-level metadata of FSD50K-Solo alongside this paper to support reproducibility and further investigation. Additionally, we release metadata generated by splitting each file into fixed-length chunks using a sliding window and providing predictions for each chunk.
Finally, in Section~\ref{sec:conclusion}, we discuss broader implications and applications of clean single-source data for advancing audio deep learning research.

\section{Related Work}
\label{sec:related_work}

Our work spans several research areas in audio machine learning: dataset quality assessment, automated data filtering methods, and pre-trained audio representation learning. While significant progress has been made in developing neural architectures for audio tasks, the fundamental challenge of ensuring high-quality training data remains underexplored. This section reviews the current state of these areas, existing limitations in large-scale audio datasets, and the role of pre-trained encoders in addressing data quality challenges.


\subsection{Quality Issues in Large-Scale Audio Datasets}
\label{ssec:problems_fsd}
Large-scale audio datasets from crowd-sourced platforms often suffer from quality degradations that can compromise model training~\cite{shawn2021}. FSD50K~\cite{fonseca2021} is subject to these challenges with three main issues: (i) high-level background noise interfering with labeled events, (ii) overlapping sound events, and (iii) temporal sparsity where target events constitute small fractions of recordings.

FSD50K's metadata includes crowd-sourced human ratings, which could theoretically be useful for filtering certain problematic data. A rating of Present and Predominant (PP) for a given class indicates ``there are no other types of sound.", and previous works such as Veluri et al.~\cite{veluri2023semantic} use this as their data filtering criteria. However, as the authors of FSD50K state, these ratings are subjective and prone to human error, especially because the ratings are crowd-sourced. For example, \textit{418989.wav} in the dev set is unanimously ranked as PP for the class \textit{Conversation}, but contains both speech and music. Although \textit{Music} is then included in the more exhaustive ground-truth labels, this complicates the curation of a single-source dataset,  which could be essential for tasks requiring isolated events such as such as source separation, event localization, and controllable audio synthesis. Conversely, filtering on human ratings can also lead to discarding useful data. \textit{347626.wav} contains a harp pluck, but the raters are unable to reach consensus on the presence of the class. Moreover, human annotation can be prohibitively costly for larger corpora, motivating the need for scalable automatic data curation.

\subsection{Automated Data Filtering}
\label{ssec:related_filtering}

Automated data filtering approaches in the audio domain have primarily focused on speech corpora, with limited exploration of general sound events. In the speech community, LibriTTS~\cite{zen19_interspeech}, a high-quality derivative of LibriSpeech~\cite{panayotov2015}, uses waveform amplitude distribution analysis (WADA)~\cite{kim08e_interspeech} to systematically exclude samples with low signal-to-noise ratios. While effective for speech, it relies on statistical properties specific to human vocal characteristics and is not directly transferable to the diverse acoustic properties of environmental sound events.
Elsewhere, deep learning-based perceptual speech quality assessment models, such as DNSMOS~\cite{reddy2022dnsmos}, have also been employed to filter speech data~\cite{ravenscroft25_interspeech}.

Recent efforts in multimodal audio datasets have adopted alternative strategies. WavCaps~\cite{mei2024} aggregates multiple audio collections and leverages ChatGPT to generate coherent and descriptive captions for each audio sample. However, audio signal-based filtering remained rudimentary, applying only duration-based filtering to exclude very short examples without analyzing the actual audio content or acoustic quality.

Audiobox Aesthetics~\cite{tjandra2025metaaudioboxaestheticsunified} is a deep learning-based perceptual audio quality prediction model for assessment of speech, music, and sound events. It predicts ratings across four dimensions: Content Enjoyment (CE), Content Usefulness (CU), Production Complexity (PC), and Production Quality (PQ). By definition, PC and PQ scores are relevant for assessing general sound event datasets in terms of the number of sources and sound quality.
Thus, we include both scores as validation criteria to benchmark the performance of our framework from a perspective different from classification metrics.



\subsection{Pre-trained Audio Encoders}
\label{ssec:related_encoders}

Large-scale pre-trained audio encoders provide powerful feature extractors that capture both acoustic and semantic properties useful for downstream tasks. On one hand, multimodal contrastive learning models have proven effective in tasks such as classification and retrieval. The Contrastive Language-Audio Pre-training (CLAP) model~\cite{elizalde2023clap} is one popular example. On the other hand, self-supervised learning (SSL) encoders have been widely adopted in audio, such as HuBERT~\cite{hsu2021hubert}, WavLM~\cite{chen2022wavlm}, and BEATs~\cite{pmlr-v202-chen23ag}. Notably, 
BEATs is trained specifically on AudioSet using masked audio modeling, enabling it to bridge the gap between speech and general audio understanding.

\section{Methods}
\label{sec:methods}


\subsection{Training Dataset Creation}
\label{ssec:data generation}

 To create the training dataset for our model, we adopt the following pipeline: (i) curating clean, single-source audio samples, 
 (ii) generating noisy mixtures with controlled criteria, and (iii) preprocessing and augmentation.

\subsubsection{Single-source Audio Generation}
Obtaining a corpus of clean and single-source audio samples with a sufficient variety to train and evaluate our system is difficult due to issues described in Sections~\ref{sec:intro} and~\ref{sec:related_work}. We explore leveraging recent advances in generative audio models for this purpose by prompting them in such a way to ensure clean single-source audio.

We use FSD50K's class labels as the target classes. First, we split its class labels into two groups based on class descriptions:
(i) \textit{single-source classes}, defined by the presence of a single, isolated sound source (e.g., barking, siren, glass breaking), and  
(ii) \textit{complex-scene classes}, defined by the presence of multiple concurrent sound events (e.g., human group activities, subway, music).

We utilize only the single-source classes, since generating audio based on complex-scene labels may result in overlapping sources or interference. 
For each selected class, 
we use Stable Audio Open 1.0 \cite{evans2024stableaudioopen} with prompts $P^{+}$ = ``The sound of \textless class\textgreater\ without any noise" and $P^{-}$ = ``Bad quality" to generate clean, single-source audio. All generated clips are manually inspected to remove those with noise. We generate 105 single-event classes with 30 samples per class (20 seconds each) at 16 kHz, serving as reference signals for training.

\subsubsection{Mixture Generation}

To generate mixtures, we first select target segments using a sliding-window maximum-energy method. Each file is chunked with window length $L$, randomly selected within $(1,10)$ seconds. For each mixture, the target segment is randomly selected from the top-5 highest energy chunks of the file.

Multi-source samples are generated by mixing the selected single-source target segment
with additional signals under four conditions with equal probability: (i) single interference, (ii) dual interference, (iii) background noise from the TAU Urban Acoustic Scenes dataset~\cite{heittola_2022_tau}, and (iv) interference plus background noise. The SNR values are uniformly sampled between $-10$ dB and $+15$ dB. To avoid semantic overlap, interference audio is not selected from classes that are highly similar to the target class (e.g., bell vs. cowbell, liquid vs. water). The dataset is constructed to enforce a 1:1 ratio between single-source and multi-source samples.
Contrary to the single-source generated samples, the multi-source samples for training and testing are not manually inspected.

\subsubsection{Pre-processing and Augmentation}
\label{sssec:augmentation}
All samples undergo root-mean-square (RMS) normalization with a target level of -26 dBFS, 
to mitigate bias from amplitude variations. 
Leading silence in each audio file is removed. To further enhance model robustness, given that real-world sound events can be repetitive, we apply temporal repetition augmentation during training where each audio sample is repeated a random number of times, uniformly sampled from $1$ to $4$, with probability $p = 0.5$. The maximum length of audio during training is 10 seconds.

\subsection{Classifier Model Design}
\label{ssec:model design}
\begin{figure}[!tb]
  \centering
  \includegraphics[width=0.80\linewidth]{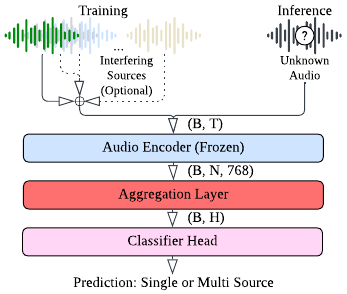}
  \vspace{-1.0mm}
  \caption{Overview of the proposed system}
  \label{fig:architecture}
  \vspace{-4.0mm}
\end{figure}

The model architecture is illustrated in Fig.~\ref{fig:architecture}. We choose the BEATs~\cite{pmlr-v202-chen23ag} encoder, with the implementation and weights from UniLM repository\footnote{Link: https://github.com/microsoft/unilm/tree/master/beats}. For temporal aggregation, we use a bidirectional long short-term memory (Bi-LSTM) network, which is used widely for audio embedding networks~\cite{wan2018_icassp_ge2e}, with a single layer with hidden size $512$. The MLP following the Bi-LSTM consists of one $1024 \times 512$ dense layer, followed by ReLU and dropout, and then a $512 \times 1$ dense layer.

\section{Experiments}
\label{sec:experiments}


\subsection{Training Details}
\label{ssec:training details}
Following the procedure described in Section~\ref{ssec:data generation}, we construct training, validation, and test sets from the generated dataset with a ratio of 8:1:1, applying data augmentation only during training. 
We use the AdamW opmitizer with a learning rate of $10^{-4}$ and weight decay of $0.01$. Training is conducted for 20 epochs with a cosine learning rate schedule with 10\% warm-up. The loss is binary cross-entropy, and the model from the epoch with the highest validation accuracy is selected. 

\subsection{Evaluation Datasets}
\label{ssec:evaluation dataset}

We acknowledge that training and evaluating on generated data presents potential domain gap concerns, where models trained on generated data may not generalize to real data. However, since large-scale, human labeled single-source versus multi-source reference datasets for general sound events do not exist, we face fundamental limitations in evaluation. Given these constraints, in addition to the generated data, we employ an internal Bose Sound Events (BSE) dataset composed of around 20 hours of expert-curated single-source to multi-source pairs, as our held-out performance benchmark.


\begin{figure*}[t]
\centering
\resizebox{2\columnwidth}{!}{
  \centering
  \includegraphics[width=1.0\linewidth]{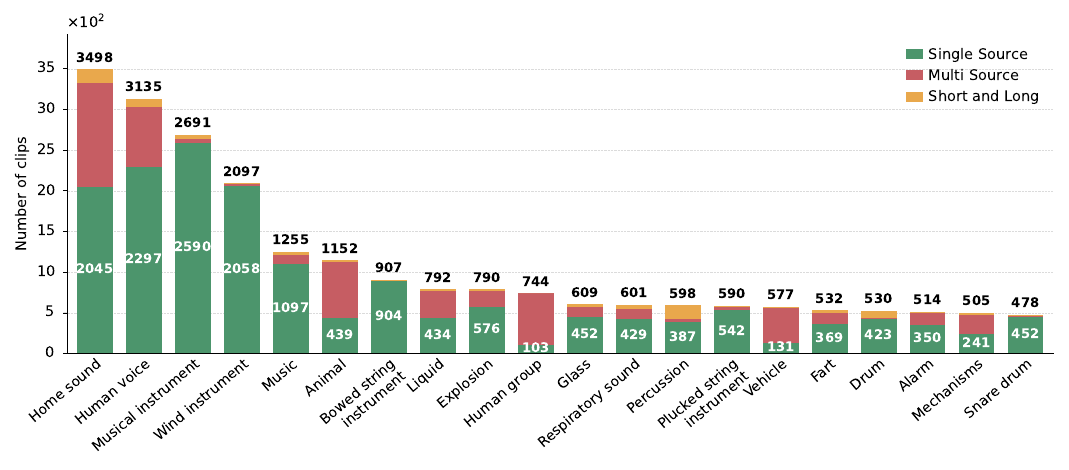}
}
\vspace{-1.em}
\caption{Top 20 Classes of FSD50K-dev. Note that ``Short and Long" illusrates the removed portion. Numbers in white is the total count of Single Source samples of each class, numbers on top of each class denote the total class sample size.}
\label{fig:top_20_classes}
\end{figure*}

\subsection{Evaluation Metrics}
\label{ssec:evaluation metrics}

For evaluation, we first employ traditional classification metrics such as accuracy, precision, recall, and F1-score. Going beyond traditional metrics, we utilize the Audiobox Aesthetics predictor~\cite{tjandra2025metaaudioboxaestheticsunified} to compute PC and PQ scores. The PC scores range from 1 to 10, where lower values imply fewer sound sources and less complexity. The PQ scores also range from 1 to 10, where higher values imply better audio quality. These metrics provide objective measures based on statistical models of human audio perception to further demonstrate the effectiveness of our automatic data curation.

\section{Results and Discussion}
\label{sec:results}

\subsection{Classification Performance}
\label{ssec:classification}

The model is evaluated on both generated and BSE test sets as shown in Table~\ref{tab:classification_performance}, where it demonstrates strong classification performance. 
Notably, the model performs better on BSE data with 2.04 percentage points higher accuracy (95.51\% vs 93.47\%) and significantly higher precision (98.58\% vs 89.31\%), demonstrating effective generalization to practical audio data filtering scenarios.

\renewcommand{\arraystretch}{1.00}
\begin{table}[h]
\vspace{-2.0mm} 
\centering
\caption{Classification performance on generated and BSE test sets.}
\begin{tabular}{l c c c c}
\toprule
\textbf{Test Set} & \textbf{Prec. $\uparrow$} & \textbf{Rec. $\uparrow$} & \textbf{F1 $\uparrow$} & \textbf{Acc. $\uparrow$} \\
\midrule
\textbf{Generated} & 89.31\% & 98.72\% & 93.81\% & 93.47\% \\
\textbf{BSE} & 98.58\% & 92.35\% & 95.36\% & 95.51\% \\
\bottomrule
\end{tabular}
\label{tab:classification_performance}
\vspace{-2.0mm} 
\end{table}

\subsection{Statistics of FSD50K-Solo}

\renewcommand{\arraystretch}{1.05}
\begin{table}[]
\vspace{-2mm} 
\centering
\caption{Statistics and quality metrics across test sets and FSD50K-Solo dataset splits. 
SS: Single-source, MS: Multi-source, Prop.: Proportion}
\resizebox{\columnwidth}{!}{ 
\begin{tabular}{@{}l@{\hspace{0.8em}}l@{\hspace{0.8em}}c@{\hspace{0.8em}}c@{\hspace{0.8em}}c@{\hspace{0.8em}}c@{}}
\toprule[1pt]
\textbf{Test Dataset} & \textbf{Label} & \textbf{Samples} & \textbf{Prop.} & \textbf{PC $\pm 2\sigma$ $\downarrow$} & \textbf{PQ $\pm 2\sigma$ $\uparrow$} \\
\midrule[1pt]
\multirow{2}{*}{\textbf{Generated}} & SS & $702$ & $50.0\%$ & $\mathbf{2.05}\pm0.70$ & $\mathbf{6.85}\pm1.81$ \\
& MS & $702$ & $50.0\%$ & $2.98\pm1.44$ & $6.26\pm1.89$ \\
\cmidrule(lr){1-6}
\multirow{2}{*}{\textbf{BSE}} & SS & $22{,}400$ & $50.0\%$ & $\mathbf{1.93}\pm0.63$ & $\mathbf{7.33}\pm1.41$ \\
& MS & $22{,}400$ & $50.0\%$ & $3.31\pm2.12$ & $6.06\pm2.21$ \\
\midrule[1pt]
\textbf{Filtered Split} & \textbf{Label} & \textbf{Samples} & \textbf{Prop.} & \textbf{PC $\pm 2\sigma$ $\downarrow$} & \textbf{PQ $\pm 2\sigma$ $\uparrow$} \\
\midrule[1pt]
\multirow{2}{*}{\textbf{FSD50K-Solo Dev}} & SS & $27{,}273$ & $69.17\%$ & $\mathbf{2.02}\pm0.89$ & $\mathbf{6.90}\pm1.60$ \\
& MS & $12{,}155$ & $30.83\%$ & $2.87\pm1.56$ & $6.41\pm1.69$ \\
\cmidrule(lr){1-6}
\multirow{2}{*}{\textbf{FSD50K-Solo Eval}} & SS & $5{,}607$ & $55.91\%$ & $\mathbf{2.14}\pm1.07$ & $\mathbf{6.92}\pm1.64$ \\
& MS & $4{,}422$ & $44.09\%$ & $2.84\pm1.61$ & $6.49\pm1.68$ \\
\bottomrule[1pt]
\end{tabular}
}
\vspace{-2mm} 
\label{tab:all_stats}
\end{table}

We applied our model to create FSD50K-Solo. We exclude very short samples ($<0.5s$, 1,727 samples) and very long samples ($>30s$, 13 samples). 
Table~\ref{tab:all_stats} presents comprehensive statistics across test sets and FSD50K-Solo dataset splits. Our model classifies 69.17\% of the development set and 55.91\% of the evaluation set as single-source. Importantly, the diffusion-generated and BSE test sets exhibit consistent PC and PQ score patterns with FSD50K-Solo, where single-source samples consistently achieve lower PC and higher PQ scores. This consistency across synthetic, curated, and real-world datasets suggests that our framework generalizes effectively to practical audio data filtering scenarios.

Fig.~\ref{fig:top_20_classes} shows the distribution of our model predictions for the top 20 classes in the FSD50K development set. For multi-label samples, the label is determined by selecting the class with the highest CLAP similarity to the audio signal. The top 5 classes stay the same, while ``Animal", ``Liquid" and ``Human group actions" are greatly reduced, causing them to fall out of the top 10 after applying our cleaning method. This is expected because many, if not most, sounds in these categories are multi-source by definition.

\begin{figure}[hb]
\vspace{-1.5mm}
\resizebox{\columnwidth}{!}{
  \centering
  \includegraphics{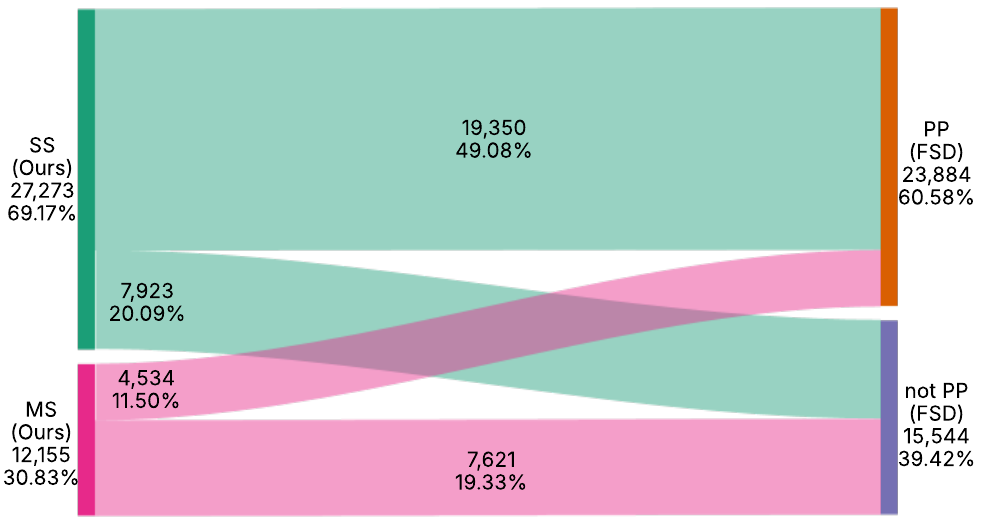}
}
\caption{Alluvial visualization of annotation flow between our model predictions and FSD50K-dev human ratings.}
\label{fig:pp_pnp_comparison}
\vspace{-1.5mm}
\end{figure}

In Fig.~\ref{fig:pp_pnp_comparison}, we compare FSD50K-Solo against FSD50K human ratings for the development set, where we accept a sample if at least two annotators rate the given class as PP.
After excluding short and long samples, filtering with PP annotations preserves 60.58\% of the data, whereas our method preserves 69.17\%. 



\section{Conclusion}
\label{sec:conclusion}


This paper presents a systematic framework for automatically identifying high-quality single-source samples in large-scale audio corpora, addressing critical data quality challenges in audio machine learning.
Our approach achieves 95.51\% accuracy on the human-curated BSE test set and filters FSD50K to create FSD50K-Solo, which contains 32,880 samples with significantly improved quality scores.
By releasing our methodology and FSD50K-Solo metadata, we establish a scalable paradigm for audio dataset curation that extends beyond FSD50K to other large-scale corpora. While our framework demonstrates strong performance on the evaluated sound classes, the model's ability to generalize to unseen event classes remains unexplored. Future work should investigate cross-domain transferability and explore zero-shot approaches.



{\normalsize\bibliographystyle{IEEEtran}
\bibliography{refs}}

@INPROCEEDINGS{10094617,
  author={Weck, Benno and Serra, Xavier},
  booktitle={ICASSP}, 
  title={Data Leakage in Cross-Modal Retrieval Training: A Case Study}, 
  year={2023},
  pages={1-5},
  doi={10.1109/ICASSP49357.2023.10094617}}

@INPROCEEDINGS{shawn2021,
  author={Hershey, Shawn and Ellis, Daniel P W and Fonseca, Eduardo and Jansen, Aren and Liu, Caroline and Channing Moore, R and Plakal, Manoj},
  booktitle={ICASSP}, 
  title={The Benefit of Temporally-Strong Labels in Audio Event Classification}, 
  year={2021},
  pages={366-370},
  doi={10.1109/ICASSP39728.2021.9414579}}

@ARTICLE{mei2024,
  author={Mei, Xinhao and Meng, Chutong and Liu, Haohe and Kong, Qiuqiang and Ko, Tom and Zhao, Chengqi and Plumbley, Mark D. and Zou, Yuexian and Wang, Wenwu},
  journal={TASLP}, 
  title={WavCaps: A ChatGPT-Assisted Weakly-Labelled Audio Captioning Dataset for Audio-Language Multimodal Research}, 
  year={2024},
  volume={32},
  pages={3339-3354},
  doi={10.1109/TASLP.2024.3419446}}

@article{fonseca2021,
author = {Fonseca, Eduardo and Favory, Xavier and Pons, Jordi and Font, Frederic and Serra, Xavier},
title = {FSD50K: An Open Dataset of Human-Labeled Sound Events},
year = {2021},
volume = {30},
journal = {TASLP},
pages = {829–852},
doi = {10.1109/TASLP.2021.3133208}
}

@misc{tjandra2025metaaudioboxaestheticsunified,
      title={Meta Audiobox Aesthetics: Unified Automatic Quality Assessment for Speech, Music, and Sound}, 
      author={Andros Tjandra and Yi-Chiao Wu and Baishan Guo and John Hoffman and Brian Ellis and Apoorv Vyas and Bowen Shi and Sanyuan Chen and Matt Le and Nick Zacharov and Carleigh Wood and Ann Lee and Wei-Ning Hsu},
      year={2025},
      eprint={2502.05139},
      archivePrefix={arXiv},
      primaryClass={cs.SD},
      url={https://arxiv.org/abs/2502.05139}, 
}

@ARTICLE{chen2022wavlm,
  author={Chen, Sanyuan and Wang, Chengyi and Chen, Zhengyang and Wu, Yu and Liu, Shujie and Chen, Zhuo and Li, Jinyu and Kanda, Naoyuki and Yoshioka, Takuya and Xiao, Xiong and Wu, Jian and Zhou, Long and Ren, Shuo and Qian, Yanmin and Qian, Yao and Wu, Jian and Zeng, Michael and Yu, Xiangzhan and Wei, Furu},
  journal={JSTSP}, 
  title={WavLM: Large-Scale Self-Supervised Pre-Training for Full Stack Speech Processing}, 
  year={2022},
  volume={16},
  number={6},
  pages={1505-1518},
  doi={10.1109/JSTSP.2022.3188113}}

@InProceedings{pmlr-v202-chen23ag,
  title = {{BEAT}s: Audio Pre-Training with Acoustic Tokenizers},
  author = {Chen, Sanyuan and Wu, Yu and Wang, Chengyi and Liu, Shujie and Tompkins, Daniel and Chen, Zhuo and Che, Wanxiang and Yu, Xiangzhan and Wei, Furu},
  booktitle = {ICML},
  pages = {5178--5193},
  year = {2023},
  volume = {202},
  publisher = {PMLR}
}

@ARTICLE{hsu2021hubert,
  author={Hsu, Wei-Ning and Bolte, Benjamin and Tsai, Yao-Hung Hubert and Lakhotia, Kushal and Salakhutdinov, Ruslan and Mohamed, Abdelrahman},
  journal={TASLP}, 
  title={HuBERT: Self-Supervised Speech Representation Learning by Masked Prediction of Hidden Units}, 
  year={2021},
  volume={29},
  pages={3451-3460},
  doi={10.1109/TASLP.2021.3122291}}

@INPROCEEDINGS{elizalde2023clap,
  author={Elizalde, Benjamin and Deshmukh, Soham and Ismail, Mahmoud Al and Wang, Huaming},
  booktitle={ICASSP}, 
  title={CLAP Learning Audio Concepts from Natural Language Supervision}, 
  year={2023},
  pages={1-5},
  doi={10.1109/ICASSP49357.2023.10095889}}

@inproceedings{zen19_interspeech,
  title = {LibriTTS: A Corpus Derived from LibriSpeech for Text-to-Speech},
  author = {Heiga Zen and Viet Dang and Rob Clark and Yu Zhang and Ron J. Weiss and Ye Jia and Zhifeng Chen and Yonghui Wu},
  year = {2019},
  booktitle = {Interspeech},
  pages = {1526--1530}
}

@inproceedings{kim08e_interspeech,
  title = {Robust signal-to-noise ratio estimation based on waveform amplitude distribution analysis},
  author = {Chanwoo Kim and Richard M. Stern},
  year = {2008},
  booktitle = {Interspeech},
  pages = {2598--2601}
}

@INPROCEEDINGS{panayotov2015,
  author={Panayotov, Vassil and Chen, Guoguo and Povey, Daniel and Khudanpur, Sanjeev},
  booktitle={ICASSP}, 
  title={Librispeech: An ASR corpus based on public domain audio books}, 
  year={2015},
  pages={5206-5210},
  doi={10.1109/ICASSP.2015.7178964}}

@misc{goel2025audioflamingo3advancing,
      title={Audio Flamingo 3: Advancing Audio Intelligence with Fully Open Large Audio Language Models}, 
      author={Arushi Goel and Sreyan Ghosh and Jaehyeon Kim and Sonal Kumar and Zhifeng Kong and Sang-gil Lee and Chao-Han Huck Yang and Ramani Duraiswami and Dinesh Manocha and Rafael Valle and Bryan Catanzaro},
      year={2025},
      eprint={2507.08128},
      archivePrefix={arXiv},
      primaryClass={cs.SD},
      url={https://arxiv.org/abs/2507.08128}, 
}

@inproceedings{
tang2024salmonn,
title={{SALMONN}: Towards Generic Hearing Abilities for Large Language Models},
author={Changli Tang and Wenyi Yu and Guangzhi Sun and Xianzhao Chen and Tian Tan and Wei Li and Lu Lu and Zejun MA and Chao Zhang},
booktitle={ICLR},
year={2024},
url={https://openreview.net/forum?id=14rn7HpKVk}
}

@misc{xu2025qwen25omnitechnicalreport,
      title={Qwen2.5-Omni Technical Report}, 
      author={Jin Xu and Zhifang Guo and Jinzheng He and Hangrui Hu and Ting He and Shuai Bai and Keqin Chen and Jialin Wang and Yang Fan and Kai Dang and Bin Zhang and Xiong Wang and Yunfei Chu and Junyang Lin},
      year={2025},
      eprint={2503.20215},
      archivePrefix={arXiv},
      primaryClass={cs.CL},
      url={https://arxiv.org/abs/2503.20215}, 
}

@ARTICLE{hu2025pseld,
  author={Hu, Jinbo and Cao, Yin and Wu, Ming and Kang, Fang and Yang, Feiran and Wang, Wenwu and Plumbley, Mark D. and Yang, Jun},
  journal={TASLP}, 
  title={PSELDNets: Pre-Trained Neural Networks on a Large-Scale Synthetic Dataset for Sound Event Localization and Detection}, 
  year={2025},
  volume={33},
  pages={2845-2860},
  doi={10.1109/TASLPRO.2025.3587446}}

@INPROCEEDINGS{shashaank2023hiss,
  author={Shashaank, N and Banar, Berker and Izadi, Mohammad Rasool and Kemmerer, Jeremy and Zhang, Shuo and Huang, Chuan-Che Jeff},
  booktitle={ICASSP}, 
  title={HiSSNet: Sound Event Detection and Speaker Identification via Hierarchical Prototypical Networks for Low-Resource Headphones}, 
  year={2023},
  pages={1-5},
  doi={10.1109/ICASSP49357.2023.10094788}}

@ARTICLE{labbe2024,
  author={Labb{\'e}, {\'E}tienne and Pellegrini, Thomas and Pinquier, Julien},
  journal={TASLP}, 
  title={CoNeTTE: An Efficient Audio Captioning System Leveraging Multiple Datasets With Task Embedding}, 
  year={2024},
  volume={32},
  pages={3785-3794},
  doi={10.1109/TASLP.2024.3430813}}

@inproceedings{veluri2023semantic,
author = {Veluri, Bandhav and Itani, Malek and Chan, Justin and Yoshioka, Takuya and Gollakota, Shyamnath},
title = {Semantic Hearing: Programming Acoustic Scenes with Binaural Hearables},
year = {2023},
booktitle = {Proc. ACM UIST},
pages = {89:1--89:15},
doi = {10.1145/3586183.3606779}
}

@INPROCEEDINGS{veluri2023realtime,
  author={Veluri, Bandhav and Chan, Justin and Itani, Malek and Chen, Tuochao and Yoshioka, Takuya and Gollakota, Shyamnath},
  booktitle={ICASSP}, 
  title={Real-Time Target Sound Extraction}, 
  year={2023},
  pages={1-5},
  doi={10.1109/ICASSP49357.2023.10094573}}

@inproceedings{wakayama25_interspeech,
  title = {{Real-time TSE demonstration via SoundBeam with KD}},
  author = {Keigo Wakayama and Tomoko Kawase and Takafumi Moriya and Marc Delcroix and Hiroshi Sato and Tsubasa Ochiai and Masahiro Yasuda and Shoko Araki},
  year = {2025},
  booktitle = {Interspeech},
  pages = {3529--3530}
}

@misc{evans2024stableaudioopen,
      title={Stable Audio Open}, 
      author={Zach Evans and Julian D. Parker and CJ Carr and Zack Zukowski and Josiah Taylor and Jordi Pons},
      year={2024},
      eprint={2407.14358},
      archivePrefix={arXiv},
      primaryClass={cs.SD},
      url={https://arxiv.org/abs/2407.14358}, 
}

@dataset{heittola_2022_tau,
  author       = {Heittola, Toni and
                  Mesaros, Annamaria and
                  Virtanen, Tuomas},
  title        = {TAU Urban Acoustic Scenes 2022 Mobile, Development
                   dataset
                  },
  month        = mar,
  year         = 2022,
  publisher    = {Zenodo},
  doi          = {10.5281/zenodo.6337421},
  url          = {https://doi.org/10.5281/zenodo.6337421},
}

@inproceedings{wan2018_icassp_ge2e,
author = {Wan, Li and Wang, Quan and Papir, Alan and Moreno, Ignacio Lopez},
title = {Generalized End-to-End Loss for Speaker Verification},
year = {2018},
url = {https://doi.org/10.1109/ICASSP.2018.8462665},
doi = {10.1109/ICASSP.2018.8462665},
booktitle = {ICASSP},
pages = {4879–4883},
numpages = {5},
}

@INPROCEEDINGS{reddy2022dnsmos,
  author={Reddy, Chandan K A and Gopal, Vishak and Cutler, Ross},
  booktitle={ICASSP}, 
  title={Dnsmos P.835: A Non-Intrusive Perceptual Objective Speech Quality Metric to Evaluate Noise Suppressors}, 
  year={2022},
  volume={},
  number={},
  pages={886-890},
  doi={10.1109/ICASSP43922.2022.9746108}}

@INPROCEEDINGS{gemmeke2017audio,
  author={Gemmeke, Jort F. and Ellis, Daniel P. W. and Freedman, Dylan and Jansen, Aren and Lawrence, Wade and Moore, R. Channing and Plakal, Manoj and Ritter, Marvin},
  booktitle={ICASSP}, 
  title={Audio Set: An ontology and human-labeled dataset for audio events}, 
  year={2017},
  volume={},
  number={},
  pages={776-780},
  doi={10.1109/ICASSP.2017.7952261}}

@inproceedings{li2025moredatacurationmatters,
      title={Less is More: Data Curation Matters in Scaling Speech Enhancement}, 
      author={Chenda Li and Wangyou Zhang and Wei Wang and Robin Scheibler and Kohei Saijo and Samuele Cornell and Yihui Fu and Marvin Sach and Zhaoheng Ni and Anurag Kumar and Tim Fingscheidt and Shinji Watanabe and Yanmin Qian},
      year={2025},
      booktitle={ASRU},
      eprint={2506.23859},
      archivePrefix={arXiv},
      primaryClass={eess.AS},
      url={https://arxiv.org/abs/2506.23859}, 
}

@inproceedings{ravenscroft25_interspeech,
  title     = {{Whilter: A Whisper-based Data Filter for "In-the-Wild" Speech Corpora Using Utterance-level Multi-Task Classification }},
  author    = {William Ravenscroft and George Close and Kit Bower-Morris and Jamie Stacey and Dmitry Sityaev and Kris Y. Hong},
  year      = {2025},
  booktitle = {{Interspeech 2025}},
  pages     = {4288--4292},
  doi       = {10.21437/Interspeech.2025-160},
  issn      = {2958-1796},
}

@inproceedings{dong2018musegan,
  title={Musegan: Multi-track sequential generative adversarial networks for symbolic music generation and accompaniment},
  author={Dong, Hao-Wen and Hsiao, Wen-Yi and Yang, Li-Chia and Yang, Yi-Hsuan},
  booktitle={Proceedings of the AAAI conference on artificial intelligence},
  volume={32},
  number={1},
  year={2018}
}

@article{ma2025real,
  title={Real-time audio-visual speech enhancement using pre-trained visual representations},
  author={Ma, T Aleksandra and Yin, Sile and Yang, Li-Chia and Zhang, Shuo},
  journal={arXiv preprint arXiv:2507.21448},
  year={2025}
}

@article{yang2026survey,
  title={A Survey of Advancing Audio Super-Resolution and Bandwidth Extension from Discriminative to Generative Models},
  author={Yang, Ningyuan and Li, Yize and Cuji, Diego A and Corey, Ryan M and Zhao, Pu and Lin, Xue and Singer, Andrew C},
  journal={arXiv preprint arXiv:2605.16681},
  year={2026}
}

@inproceedings{luo2018tasnet,
  title={Tasnet: time-domain audio separation network for real-time, single-channel speech separation},
  author={Luo, Yi and Mesgarani, Nima},
  booktitle={2018 IEEE International Conference on Acoustics, Speech and Signal Processing (ICASSP)},
  pages={696--700},
  year={2018},
  organization={IEEE}
}

@inproceedings{irvin2024fullband,
  title={A fullband neural network for audio packet loss concealment},
  author={Irvin, Bryce and Yin, Sile and Zhang, Shuo and Stamenovic, Marko},
  booktitle={2024 IEEE International Conference on Acoustics, Speech, and Signal Processing Workshops (ICASSPW)},
  pages={107--108},
  year={2024},
  organization={IEEE}
}

\end{document}